# Composable Ledgers for Distributed Synchronic Web Archiving

Thien-Nam Dinh
thidinh@sandia.gov
Sandia National Labs
Albuquerque, NM, USA

Nicholas Pattengale
ndpatte@sandia.gov
Sandia National Labs
Albuquerque, NM, USA

## ABSTRACT
The Synchronic Web is a highly scalable notary infrastructure that provides tamper-evident data provenance for historical web data. In this document, we describe the applicability of this infrastructure for web archiving across three envisioned stages of adoption. We codify the core mechanism enabling the value proposition: a procedure for splitting and merging cryptographic information fluidly across blockchain-backed ledgers. Finally, we present preliminary performance results that indicate the feasibility of our approach for modern web archiving scales.

## 1 INTRODUCTION

In the effort to preserve digital history for future generations, blockchain technology presents a compelling value proposition: the ability to enforce secure data provenance across discrete time. While researchers have demonstrated viable approaches in such diverse use cases as timestamping [4], document editing [5], and academic submissions [3], large-scale adoption remains elusive. We suggest that the recent codification of the Synchronic Web [2], a characteristically simple, scalable, and generalizable blockchain infrastructure, maybe the technical innovation needed to achieve critical mass in this domain. Such an infrastructure, once fully realized, would provide the ability for entities around the world to cryptographically prove and verify the provenance of domain-agnostic digital content—creating a foundational notion of credibility that is achievable by all compliant archiving entities. Our work advances this endeavor by identifying and codifying a key procedure in this paradigm: the decomposition and recomposition of Synchronic Web commitments[1] needed to securely move data between archives. We model the former as a *split* operation and the latter as a *merge* operation on the local ledger containing the cryptographic metadata. Through this work, we establish the possibility of secure and fluid data movement within the dynamic web archiving ecosystem.

## 2 USAGE

This section describes the importance of the split/merge procedure across three envisioned stages of adoption.

---
[1]A commitment is a piece of metadata asserting the provenance of a piece of content.

### 2.1 Data Management

The first stage considers adoption by only a single web archive organization. If the organization operates a distributed, large-scale, or otherwise complex architecture [1], then it could benefit from the inexpensive and flexible provenance provided by the Synchronic Web. In this stage, the split/merge procedure is necessary for persisting historical commitments through infrastructural changes. As the organization makes changes to either its data management or digital identity architecture, the procedure would allow it to migrate the original commitments without loss of security.

### 2.2 Data Sharing

The second stage considers adoption by multiple web archive organizations. Since different archives preserve different types of content, a collection of archives would benefit from the characteristically interoperable and domain-agnostic provenance provided by the Synchronic Web. In this stage, the function of the split/merge procedure is to securely transfer previously collected data from one archive to another. The need for such data transfers may arise, for example, when old archives are discontinued, when new archives emerge, or when two archives have overlapping collection interests.

### 2.3 Data Provenance

The third stage consists of adoption by non-archival organizations. For a variety of reasons[2], normal (non-archival) websites may wish to secure their content with Synchronic Web commitments. In this stage, websites would split off commitments to display on their website such that they can be collected and merged into web archives. Figure 1 provides a visual of the envisioned workflow.

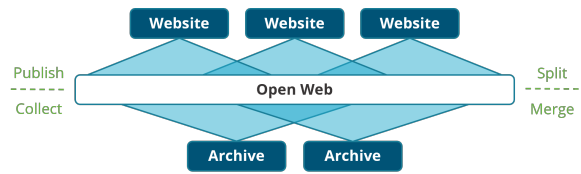

Figure 1: Provenance Flow. Commitments are split by websites for display on the open web and merged by archives.

## 3 DESIGN

This section codifies the split/merge procedure within the standard Synchronic Web setup[3]. In the standard setup, archivable *content* is backed by ledgers of *commitments* that are secured by *notaries*, maintained by *journals*, and checked by *verifiers*. Discrete time is defined by the *index* of blocks in the blockchain. Digital space is defined by a hierarchy of *verifiable maps* that each consist of a set of *keys*, *values*, and *proofs*. Given this setup, the overarching task is to transfer content from one journal to another while preserving the security guarantees of its commitments. Given this task, the core requirement is a procedure to (1) decompose an original ledger of size $n$ into a set of commitments and (2) recompose a subset of $m$ commitments into a new derivative ledger. This section describes such a procedure.

---
[2]Section 4 of the whitepaper [2] classifies several reasons.
[3]Section 3 of the whitepaper [2] provides definitions and additional details.

## 3.1 Core Procedure

Algorithm 1 defines functions to split and merge the Merkle tree that secures a verifiable map[4]. For a generic *SplitTree* operation, the input size is $O(n)$, the output transfer size is $O(m \log n)$, and the time is $O(m \log n)$. For a generic *MergeTree* operation, the input transfer size is $O(m \log n)$, the output size is $O(\frac{m \log n}{\log m})$, and the time is $O(m \log n)$. When splitting the original ledger or merging the original set of commitments, $m$ is equal to $n$.

**Algorithm 1 Split/Merge Tree.** Functions for splitting commitments from one verifiable map to and merging into another. *GetProof* is defined in Algorithm 3 of the whitepaper [2].

```
1: function SPLITTREE(map, tree)
2:     return [[key, value, GetProof(tree, key)]
3:         for each key, value ∈ map]
4: function MERGETREE(entries)
5:     function Recurse(items, depth)
6:         if items.length = 1 ∧ items[0][2].length = depth then
7:             return [Hash(items[0][0] + items[0][1]), [∅, ∅]]
8:         zeros, ones ← [ ], [ ]
9:         for each item ∈ items do
10:            if Binary(item[0])[depth] = 0 then
11:                zeros.append(item)
12:            else
13:                ones.append(item)
14:        if zeros.length = 0 then
15:            left ← Recurse(zeros, depth + 1)
16:        else
17:            left ← [ones[0][2][depth], [∅, ∅]]
18:        if ones.length = 0 then
19:            right ← Recurse(ones, depth + 1)
20:        else
21:            right ← [zeros[0][2][depth], [∅, ∅]]
22:        return Node(
23:            Hash(left[0] + right[0]), [left, right])
24:    return Recurse(entries)
```

## 3.2 Optimizations

We describe two optimizations that we leave for implementation-level design. The first is the use of *multi-proofs* [6] for transferring bulk commitments between verifiable maps. This optimization would reduce the transfer size to $O(\frac{m \log n}{\log m})$. The second is the extension of multi-proofs to compress related commitments from partially similar verifiable maps. For instance, when compressing multiple states of the same ledger across $i$ blocks, this optimization would reduce the transfer size to $O(i \frac{m\prime \log n}{\log m})$ where $m\prime$ is the subset of commitments that changes across any two contiguous steps.

---
[4]The splitting and merging of the non-cryptographic portion is considered to be trivial.

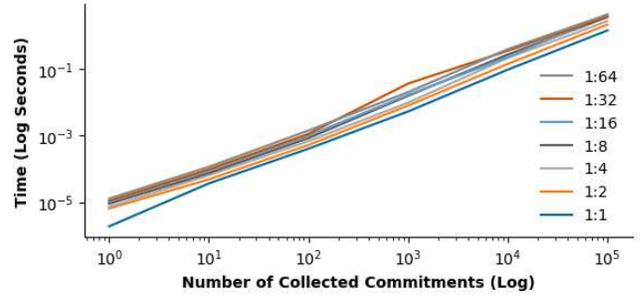

**Figure 2: MergeTree Performance.** Time elapsed after merging $x$ commitments pairs (geometric mean of 10 trials). Each series corresponds to a different ratio of $m$ to $n$.

## 4 IMPLEMENTATION

We implemented our procedure into the existing Synchronic Web prototype, which currently includes a notary server, a journal Python SDK, and a verifier browser extension. Figure 2 displays a set of experiments performed on a personal laptop with an Intel Core i9-11950H 2.60GHz CPU. These preliminary results indicate that the procedure described in this document could plausibly be deployed at the scale required by modern web archiving[5].

## 5 PATH FORWARD

The next steps are to explore concrete applications for our procedure. For web archives, work remains to integrate the Synchronic Web commitments into performant data infrastructure. On the open web, we see value in developing compatible web crawlers and browser extensions. For websites, we anticipate the emergence of a new generation of version-controlled Synchronic Web servers. Ultimately, the success of these tools will determine the success of our efforts to catalyze a new landscape for secure web archiving.

---
[5]A 2020 blog post from the Internet Archive states that the organization crawls ≈ 750 million pages per day (http://blog.archive.org/2020/12/19/looking-back-on-2020/).